%% file: MECommunityV07.tex
\documentclass[twocolumn,aps,floatfix,nofootinbib]{revtex4}

\pdfoutput=1

\usepackage[usenames,dvipsnames,svgnames]{xcolor}
\usepackage[caption=false]{subfig}
\usepackage{tikz,pgfplots,tkz-euclide}
\usepackage{float}
\usepackage{pgfplots}
\usepackage{amsmath,mathtools,amssymb}
\usepackage{tabularx,booktabs,adjustbox,multirow}
\usepackage[inline]{enumitem}

\input{Plots}

\AtBeginDocument{
\heavyrulewidth=.08em
\lightrulewidth=.05em
\cmidrulewidth=.03em
\belowrulesep=.65ex
\belowbottomsep=0pt
\aboverulesep=.4ex
\abovetopsep=0pt
\cmidrulesep=\doublerulesep
\cmidrulekern=.5em
\defaultaddspace=.5em
}

\DeclareMathOperator{\oreal}{Re}
\DeclareMathOperator{\oimag}{Im}
\newcommand{\real}[1]{\oreal\parens*{#1}}
\newcommand{\imag}[1]{\oimag\parens*{#1}}
\newcommand{\prn}[1]{\left ( #1 \right )}
\newcommand{\brq}[1]{\left [ #1 \right ]}

\newcommand{\set}[1]{\left \{ #1 \right \}}

\newcommand{\im}{\mathrm{i}}
\newcommand{\der}{\text{d}}
\newcommand{\cder}[1]{\text{D}_{#1}}
\newcommand{\inp}[1]{\langle #1 \rangle}
\newcommand{\bra}[1]{\langle #1 |}
\newcommand{\ket}[1]{| #1 \rangle}
\DeclarePairedDelimiter\pabs{\lvert}{\rvert}

\newcommand{\abs}[1]{\pabs*{#1}}

\newcommand{\iden}{\mathbb{I}}
\DeclareMathOperator{\otr}{Tr}
\newcommand{\tr}[1]{\otr\parens*{#1}}
\DeclareMathOperator{\osign}{sign}
\newcommand{\sign}[1]{\osign\parens*{#1}}
\DeclareMathOperator*{\phaseop}{phase}
\newcommand{\phase}[1]{\phaseop\parens{#1}}
\newcommand{\cst}{\text{cst}}
\newcommand{\bpsi}{\boldsymbol\psi}
\newcommand{\ws}{w_\text{s}}
\newcommand{\lap}{\mathcal{L}}
\newcommand{\laph}[1]{\hat{\lap}_{#1}}
\newcommand{\laphc}{\laph{}^{\text{C}}}
\newcommand{\laphn}[1]{\laph{#1}^\text{N}}
\newcommand{\Hh}{\hat{H}}
\newcommand{\evec}[1]{\chi_{#1}}
\newcommand{\evect}[1]{\evec{#1}^{*}}
\newcommand{\pevec}[1]{\varphi_{#1}}
\newcommand{\eval}[1]{\lambda_{#1}}
\newcommand{\Tra}[3]{T^{#1}_{#2 \to #3}}
\newcommand{\Tr}[2]{\Tra{\theta}{#1}{#2}}
\DeclarePairedDelimiter{\parens}{(}{)}
\DeclareMathOperator*{\opU}{U}
\DeclareMathOperator*{\opSO}{SO}
\DeclareMathOperator*{\opO}{O}
\newcommand{\gU}[1]{\opU\parens*{#1}}
\newcommand{\gSO}[1]{\opSO\parens*{#1}}
\newcommand{\gO}[1]{\opO\parens*{#1}}
\newcommand{\projl}{\mathbb{P}^{(\ell)}}
\newcommand{\projlh}{\hat{\mathbb{P}}^{(\ell)}}
\newcommand{\R}{\mathbb{R}}
\newcommand{\Z}{\mathbb{Z}}
\newcommand{\C}{\mathbb{C}}

\newcommand{\aH}{a_\text{H}}
\newcommand{\aM}{a_\text{M}}
\newcommand{\degs}{\deg_\text{s}}
\newcommand{\parti}[1]{P_{#1}}
\newcommand{\mH}{\mathbf{H}}
\newcommand{\dmath}[1]{\hat{\rho}_{#1}}
\newcommand{\dmat}[2]{\prn{\rho_{#1}}_{#2}}
\newcommand{\brho}[1]{\boldsymbol\rho^{\theta}_{#1}}
\newcommand{\cor}[2][]{x_{#2}^{#1}}
\newcommand{\adj}[2][]{X_{#2}^{#1}}
\newcommand{\cornt}[2][]{\xi_{#2}^{#1}}
\newcommand{\adjnt}[2][]{\Xi_{#2}^{#1}}
\newcommand{\agropt}[1]{#1}
\newcommand{\sepopt}{~}
\newcommand{\sepcon}{~}

\newcommand{\minpc}[3]{\min_{#1}\sepopt{\agropt{#2}}\sepcon\text{s.t. }#3}

\begin{document}

\title{Magnetic eigenmaps for community detection in directed networks}
\author{Micha\"el Fanuel }\email{Michael.Fanuel@esat.kuleuven.be}
\author{Carlos M. Ala\'iz}\email{CMAlaiz@esat.kuleuven.be}
\author{Johan A. K. Suykens}\email{Johan.Suykens@esat.kuleuven.be}
\affiliation{KU Leuven, Department of Electrical Engineering (ESAT)\\STADIUS Center for Dynamical Systems, Signal Processing and Data Analytics,\\Kasteelpark Arenberg 10, B-3001 Leuven, Belgium}

\date{\today}

\begin{abstract}
 Communities in directed networks have often been characterized as regions with a high density of links, or as sets of nodes with certain patterns of connection.
 Our approach for community detection combines the optimization of a quality function and a spectral clustering of a deformation of the combinatorial Laplacian, the so-called magnetic Laplacian. The eigenfunctions of the magnetic Laplacian, that we call magnetic eigenmaps, incorporate structural information. Hence, using the magnetic eigenmaps, dense communities including directed cycles can be revealed as well as ``role'' communities in networks with a running flow, usually discovered thanks to mixture models.
 Furthermore, in the spirit of the Markov stability method, an approach for studying communities at different energy levels in the network is put forward, based on a quantum mechanical system at finite temperature.
\end{abstract}

\maketitle

\section{Introduction}

The investigation of network structure has been performed with the help of a wealth of techniques~\cite{PorterNoticeAmercican} with various advantages, a famous example being modularity optimization~\cite{NewmanGirvan,NewmanPNAS} in undirected networks. One of the main disadvantages that some of these methods have to face is the resolution limit. In particular, extensions of modularity for changing the resolution limit have been developed~\cite{ScienceMucha}, whereas other related methods are inspired from statistical physics. Indeed, a general framework was developed in~\cite{Reichardt}, and later, a specific Potts model was explained to have no resolution limit~\cite{ConstantPottsModel}.
On the other hand, among the community detection methods, the discrete Laplacians in undirected networks~\cite{ChungBook,Luxburg} have also been used in order to unravel the graph structures.
A common feature of these definitions of communities is that they rely on the density of links in the network.

Recently, the community structure of complex networks has been studied with the help of several dynamical processes~\cite{DelvenneLambiotteRocha,RosvallBergstrom}. For instance, flow communities are detected in the information theory based framework ``Infomap'' of~\cite{RosvallBergstrom}. In this dynamical paradigm, the ``Markov stability'' method uses a dynamical process governed by a random walk or a continuous time Kolmogorov equation to unravel the network geometry at different scales~\cite{MarkovStabilityPNAS,MarkovStabilityIEEE}. An asset of this method is that it naturally contains  previous methods such as  modularity optimization and Fiedler partitioning. As a matter of fact, time evolution allows to span different scales, since at different times the eigenmodes of the dynamics have another relative importance. These eigenmodes incorporate structural information.

The importance of diffusion processes for investigating network structure has been highlighted in recent works~\cite{MarkovStabilityIEEE,TwitterInterestCommunities,DelvenneLambiotteRocha}. Furthermore, these approaches have emphasized the relevance of a different type of community structure: the flow-based communities, which are intuitively defined as the structures retaining the diffusion for a certain period of time.
Nonetheless, while the diffusion processes on networks are increasingly understood in the case of undirected networks, there has been less focus in the past on diffusion on directed networks. Many of the networks of interest in biology, internet or social sciences are directed and have attracted attention in the physics literature~\cite{Arenas, LeichtNewman,Kim,MalliarosVazirgiannis,Fortunato}. In dynamical frameworks, where the existence of a stationary distribution is crucial, directed networks are explored thanks to a random walk with teleportation as for instance in the Markov stability framework~\cite{MarkovStabilityIEEE}, the Infomap method~\cite{RosvallBergstrom}, or in the definition of the LinkRank method~\cite{Kim}. A natural question is: can we spare the use of a random walk with teleportation?

Although alternatives approaches to Markovian processes have also been studied in~\cite{MemoryRosvall}, we propose here a method based on quantum mechanics to uncover communities in directed networks. Our approach relies on a deformation of the combinatorial Laplacian suited to directed graphs and does not fit into the theory of Markov processes. More explicitly, the magnetic Laplacian~\cite{Berkolaiko,deVerdiere,Shubin} is a generalization of the combinatorial Laplacian to a line bundle, and can be understood as describing the dynamics of a free quantum mechanical particle on a graph under the influence of magnetic fluxes passing through the cycles in the network. It is widely known in the physics community that the presence of a magnetic flux can be detected in quantum mechanics thanks to the Aharonov-Bohm effect~\cite{AharonovBohm}.
Whereas the combinatorial Laplacian was designed long ago as a discrete differential operator, it has been used only more recently for community detection in networks~\cite{ShiMalik,ChungBook,Luxburg}. Similarly, the magnetic Laplacian is a well-known object in mathematical and condensed matter physics~\cite{deVerdiere,Berkolaiko}, however, to the best of our knowledge, it has never been used for community detection in directed networks, apart from the purely theoretical work~\cite{Lange2015}.
As a matter of fact, the major asset of our quantum mechanical approach over dynamical frameworks relying on random walks or Kolmogorov equations is that the generator of the dynamics can be here a complex valued Hermitian operator. Actually, the question of the existence of a stationary distribution is irrelevant in our case.

In Section~\ref{SecMagLap}, the properties of the magnetic Laplacian will be reviewed with a particular emphasis on its connection to the network topology. Then, the so-called flux communities will be introduced in Section~\ref{SecFluxCom}, and a method for uncovering them will be discussed in Section~\ref{SecClustering}. Subsequently, a multiscale method for studying flux communities will be proposed in Section~\ref{SecMultiScale}, while the results will be illustrated on artificial and real-life networks in Section~\ref{SecFluxARNet}. An analogue of the spectral clustering method in the complex domain will be introduced in Section~\ref{SecClusteringComplex} as a tool to uncover role communities in directed networks with a running flow. Finally, the paper will end with some conclusions in Section~\ref{SecConlusions}.

A summary of the different methods that we propose is included in Table~\ref{TabSummary}.

\begin{table*}[t]
\begin{tabular}{lccc}
\toprule
 & \multicolumn{2}{c}{Flux} & \multirow{2}{*}{Flow} \\
 \cmidrule(lr){2-3}
 & Energy $\eval{\ell}$ & Finite temperature $1/\beta$ & \\
\midrule
Illustration & \multicolumn{2}{c}{\adjustbox{valign=m}{\includetikznb{Scheme1}}} & \adjustbox{valign=m}{\includetikznb{Scheme2}} \\
 Correlation & $\cor{\ell,\theta}\prn{i,j} = \real{\evec{\theta,\ell}\prn{i} \Tr{i}{j} \evect{\theta,\ell}\prn{j}}$ & $\cor{\beta,\theta}\prn{i,j} =\real{\Tr{i}{j} \dmat{\beta}{i,j}}$ & $\cornt{\theta,0}\prn{i,j} = \real{\evec{\theta,0}\prn{i} \evect{\theta,0}\prn{j}}$ \\
 Adj.  matrix & $\adj{\ell,\theta}\prn{i,j} = \cor{\ell,\theta}\prn{i,j}+ \abs{\cor{\ell,\theta}\prn{i,j}}$ & $\adj{\beta,\theta}\prn{i,j} = \cor{\ell,\beta}\prn{i,j} + \abs{\cor{\ell,\beta}\prn{i,j}}$ & $\Xi_{\theta,0}\prn{i,j} = \cornt{\theta,0}\prn{i,j} + \abs{\xi_{\theta,0}\prn{i,j}}$ \\
 \bottomrule
 \end{tabular}
 \caption{Summary of the community detection methods using the magnetic eigenmaps $\evec{\theta,\ell}$. \label{TabSummary}}
\end{table*}

\paragraph*{Notation}

In the sequel, a directed network will be considered, with its set of nodes $V$ and links $E$. An undirected link between $i$ and $j$ will be denoted by $\set{i,j}$, while a directed link will be written $\brq{i,j}$.

\section{Magnetic Laplacian and line bundles}
\label{SecMagLap}

A conventional manner to study the structure of directed networks is to symmetrize its weight matrix in order to make the network undirected so that a spectral method, based on the combinatorial Laplacian
\begin{equation*}
 \prn{\laphc\psi} \prn{i} = \sum_{j} \ws\prn{i,j} \prn{\psi\prn{i}-\psi\prn{j}} ,
\end{equation*}
can be used to partition the network.
In general, the function of the nodes $\psi$ is taken to be real-valued. Another approach consists in defining an analogue of the PageRank random walk on the network, where the walker follows exclusively the edge directions. For technical reasons, a teleportation parameter has to be added so that a stationary distribution can be defined, if the network is not strongly connected and aperiodic. In this article, we consider an intermediate possibility which uses the symmetrization of the weights by keeping relevant information about the edge directions in an edge flow $1$-form. Indeed, decomposing the weight matrix, we define the symmetrized weight $\ws\prn{i,j} = \prn{w\prn{i,j} + w\prn{j,i}} / 2$ and the skew-symmetric non-dimensional function of the oriented links $a\prn{i,j}$, satisfying $a\prn{i,j} = 1$ if $i \to j$, $a\prn{i,j} = -1$ if $j \to i$, and $a\prn{i,j} = 0$ if $\set{i,j}$ is reciprocal. Separating the asymmetric part of the weight matrix was already proposed in~\cite{LiZhang}, however no similar deformation of the combinatorial Laplacian was proposed earlier in the literature.

\subsection{Magnetic Laplacian}

\begin{figure}[t]
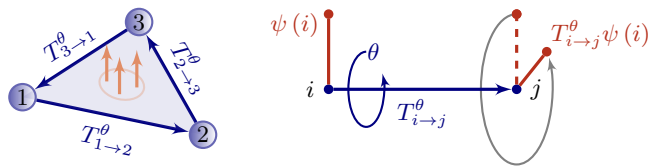

 \includetikznb{Scheme3}\hfill%
 \includetikznb{Scheme4}
 \caption{Parallel transport along a directed cycle (here a triangle is chosen as an example), and an illustration of the effect of a transport along a directed link. It is intuively clear that $\Tr{i}{j}$ assigns a rotation to a directed link. \label{FigParallelTransportTriangle}}
\end{figure}

As a consequence of these elementary remarks, we propose to describe the Hamiltonian of our quantum mechanical system as the Hamiltonian of a free charged particle on an undirected network in presence of a space varying magnetic field, given by the so-called ``magnetic Laplacian''~\cite{Berkolaiko,deVerdiere,Shubin},
\begin{equation}
 \prn{\laph{a,\im \theta} \psi}\prn{i} = \sum_{j} \ws\prn{i,j} \prn{\psi\prn{i} - \Tr{j}{i}\psi\prn{j}}, \label{MagneticLaplacian}
\end{equation}
with $\Tr{j}{i} = \exp\prn{\im \theta a(j,i)}$, depending on a real deformation parameter $\theta$ interpreted as the electric charge of a particle. Obviously, the combinatorial Laplacian of the symmetrized weight matrix is recovered if either $\theta = 0$ or $a = 0$, i.e. $\laphc = \laph{a,0} = \laph{0,\theta}$.
The dynamics is invariant under $\theta \to \theta + 2\pi$, and therefore the parameter $\theta$ is interpreted as being an angle, as illustrated in Figure~\ref{FigParallelTransportTriangle}. We may call this version of electrodynamics ``compact electrodynamics''.

The self-adjoint operator $\Hh = \laph{a, \im \theta}$ is actually positive semi-definite and can be understood as a deformation of the combinatorial Laplacian on the undirected graph. More details are given in Appendix~\ref{SecDVBL}. Furthermore, it is a special case of vector bundle Laplacian~\cite{KenyonVectorBundle,FormanDeterminant}. Indeed, the factor $\Tr{j}{i} = \exp\prn{\im \theta a(j,i)}$ is interpreted as a unitary parallel transporter. Although the magnetic Laplacian was already present in the physics literature long ago, it can be also understood in the ``connection Laplacian'' framework of Singer and Wu~\cite{SingerWu,ChungZhao}, but for the case of a complex unitary representation of $\gU{1}$ instead of real representations of $\gSO{d}$, as explained in Appendix~\ref{SecMLasCL}.
Incidentally, Cucuringu has proposed recently a ranking algorithm based on a $\gU{1}$ connection Laplacian~\cite{CucuringuSyncRank}.

\subsection{Discrete Hodge theory and gauge transformations}

The customary gauge transformation $a' = a + \der h$ of a magnetic Laplacian, for a discrete gradient $\der h\prn{i,j} = h\prn{j} - h\prn{i}$, gives a unitary equivalent operator
\begin{equation}
\label{GaugeTransformation}
 \laph{a',\im \theta} = e^{-\im \theta h} \circ \laph{a,\im \theta} \circ e^{\im \theta h} ,
\end{equation}
which obviously shares the same spectrum. Graph gauge theory has been already discussed for instance in~\cite{ChungSternberg}, where no explicit application to directed networks was presented.

The magnetic Laplacian is closely related to the topology of the network and we will take advantage of this feature in order to uncover various types of community structures. Actually, the topology of graphs and simplicial complexes has been shown to incorporate relevant information about data. Here, because electromagnetism is associated to the gauge group $\gU{1}$, the relevant topological structures uncovered are the cycles.
In this spirit, we can perform the Hodge decomposition~\cite{YaoRanking} of the edge flow
\begin{equation*}
 a = \aH + \der h_a +\der^\star \omega_a = \aM + \der h_a ,
\end{equation*}
where $\aH$ is the harmonic component related to the existence of magnetic flux through undirected $k$-cycles ($k>3$) in the network and $\der^\star \omega_a$ is a co-exact form related to a magnetic flux through the undirected triangles of the graph. This Hodge decomposition allows to consider components of the edge flow related to cycles in the network. Incidentally, these topologically non-trivial components actually play an important role in the non-perturbative dynamics of gauge theories~\cite{FanuelModular}.

An asset of our description is that the magnetic Laplacian will emphasize the importance of groups of nodes organized as a directed cycle, which is actually a structure causing problems when the directed network is explored by means of a random walk.

\section{Aharonov-Bohm effect and flux communities}
\label{SecFluxCom}

\subsection{Directed networks and Aharonov-Bohm phases}

Considering the Schr\"odinger equation on a network with the magnetic Laplacian as Hamiltonian, we can write the conservation of probabilities in terms of the divergence of a probability current. Indeed, the time evolution of the probability distribution $p_t\prn{i} = \abs{\psi(i,t)}^2$ on the network is governed by
\begin{equation}
\label{ConservationProbabilities}
 \frac{\partial}{\partial t} p_t\prn{i} = \sum_{j} \ws\prn{i,j} J_t\prn{i,j} ,
\end{equation}
with the real valued and gauge invariant probability current
\begin{equation}
\label{ImaginaryPart}
 J_t\prn{i,j} = 2 \imag{\psi(i,t) \Tr{i}{j} \psi^{*}(j,t)} ,
\end{equation}
where $\Tr{i}{j} \psi(i,t)$ is the value of $\psi$ at node $i$ and time $t$, transported at node $j$.

Although we do not have here a diffusion-like dynamics, \eqref{ConservationProbabilities} governs the time evolution of a probability. Nevertheless, we are interested in the phase correlations of the nodes and therefore we are going to study the real part in~\eqref{ImaginaryPart} instead of the imaginary part. Incidentally, since we are going to consider first the case where the state $\psi$ is an eigenstate of the magnetic Laplacian, we will not consider the associated probability current~\eqref{ImaginaryPart}.
Let us consider the orthonormal basis of functions on the nodes $\inp{\delta_i|\delta_j} = \delta_{i,j}$, localized on a specific node ($\delta_{i,j}$ is the Kr\"onecker delta).
The matrix elements of the magnetic Laplacian are
\begin{equation}
\label{CorrelationTransport}
 \inp{\delta_i|\laph{a,\im \theta}|\delta_j} = \sum_{\ell = 0}^{N-1} \eval{\ell} \evec{\theta,\ell}\prn{i} \evect{\theta,\ell}\prn{j} ,
\end{equation}
where $\evec{\theta,\ell}$ is the eigenfunction associated to the eigenvalue $\eval{\ell}$, satisfying $\eval{0} \leq \eval{1} \leq \eval{2} \leq \dots$. These eigenfunctions are called in this paper the \emph{magnetic eigenmaps}.

Actually, it is well-known that the eigenfunctions $\evec{\theta,\ell}$ are stationary states, i.e., their probability density does not evolve in time. In fact, the rhs of~\eqref{ConservationProbabilities} vanishes. Nevertheless, in general, the real part of the same matrix elements can be interesting.
Let us first assume that the eigenvalue $\eval{\ell}$ is non degenerate. Hence, we want to find the partition which maximizes the following correlation due to the magnetic field:
\begin{equation}
\label{Bilinear}
 \cor[(a)]{\ell,\theta}\prn{i,j} = \real{\evec{\theta,\ell}\prn{i}\Tr{i}{j}\evect{\theta,\ell}\prn{j}} ,
\end{equation}
which can be negative. Therefore, we introduce the following matrix with positive matrix elements:
\begin{equation}
\label{PositiveMatrixElements}
 \adj[(a)]{\theta,\ell}\prn{i,j} = \abs{\evec{\theta,\ell}\prn{i}\evect{\theta,\ell}\prn{j}} + \cor[(a)]{\theta,\ell}\prn{i,j} 
\end{equation}
for all $i,j\in V$ such that $\set{i,j}\in E$, and $\adj[(a)]{\ell,\theta}\prn{i,j} = 0$ otherwise.
The matrix elements $\adj[(a)]{\ell,\theta}\prn{i,j}$ may be understood as link weights, and they are gauge invariant, i.e. the correlation is the same either if $a = \aM +\der h$ or $\aM$ are used to compute it. Hence, we have
\begin{equation*}
\label{GaugeInvariantCurrent}
 \adj[(a)]{\theta,\ell}\prn{i,j} = \adj[(\aM)]{\theta,\ell}\prn{i,j} .
\end{equation*}
For the sake of simplicity, we will now omit the superscript indicating the dependence on $a$ and merely write $\adj{\theta,\ell} = \adj[(a)]{\theta,\ell}$. Let us explain the definition of the correlation of~\eqref{Bilinear} with the example of Figure~\ref{FigParallelTransportTriangle}.
First of all, finding the eigenvector of the magnetic Laplacian with the lowest eigenvalue can be formulated as the following optimization problems:
\begin{align}
 &{\quad} \minpc{\evec{}}{\inp{\evec{} | \laph{a,\im \theta} \evec{}}}{\inp{\evec{}|\evec{}} = 1} \nonumber \\
 &\equiv \minpc{\evec{}}{\sum_{\set{i,j} \in E} \ws\prn{i,j} \abs{\prn{\cder{a, \im \theta} \evec{}}\prn{i,j}}^2}{\inp{\evec{}|\evec{}} = 1} , \label{VariationalPrinciple}
\end{align}
where the covariant derivative is defined as
\begin{equation*}
 \prn{\cder{a,\theta}\evec{}}\prn{i,j} = e^{\im \theta a\prn{i,j} / 2} \evec{}\prn{j} - e^{- \im \theta a\prn{i,j} / 2} \evec{}\prn{i} .
\end{equation*}
Notice that the covariance property of this gradient under the transformation $a' = a + \der h$ reads
\begin{equation*}
 \prn{\cder{a',\theta}\evec{}}\prn{i,j} = e^{- \im \theta \prn{h\prn{i} + h\prn{j}} / 2} \prn{\cder{a,\theta} e^{\im \theta h}\evec{}}\prn{i,j} .
\end{equation*}

In analogy with the case of the combinatorial Laplacian, the solution to this minimization problem will satisfy $\prn{\cder{a,\im\theta}\evec{}}\prn{i,j} \approx 0$.
Let us explain why the weight $\cor[(a)]{0,\theta}\prn{i,j}$ is large if $\brq{i,j}$ is part of a directed cycle with an example.
More precisely, in the case of the directed triangle of Figure~\ref{FigParallelTransportTriangle}, if the lowest energy eigenfunction $\evec{\theta,0}$ should  satisfy $\prn{\cder{a,\im\theta}\evec{\theta, 0}}\prn{i,j} = 0$ for the three links, then
\begin{equation}
\label{EqTriangleEquations}
 \begin{cases}
  \evec{\theta,0}\prn{3} = e^{\im \theta} \evec{\theta,0}\prn{2} , \\
  \evec{\theta,0}\prn{2} = e^{\im \theta} \evec{\theta,0}\prn{1} , \\
  \evec{\theta,0}\prn{1} = e^{\im \theta} \evec{\theta,0}\prn{3} .
 \end{cases}
\end{equation}
Hence, $\exp\prn{\im 3 \theta} = 1$, so $\theta$ should be selected as $\theta = 2 \pi / 3$.
Indeed, in that case we have
\begin{align*}
 \evec{\theta,0}(1) \Tr{1}{2} \evect{\theta,0}\prn{2} &= \evec{\theta,0}(1) e^{\im \theta} \evect{\theta,0}\prn{2} \\
  &= \prn{\evec{\theta,0}(1) e^{\im \theta}} \prn{e^{\im \theta} \evec{\theta,0}\prn{1}}^{*} \\
  &= \abs{\evec{\theta,0}\prn{1}}^2 .
\end{align*}
Since this is valid for the directed links $1 \to 2$, $2 \to 3$ and $3 \to 1$, the correlations~\eqref{Bilinear} between subsequent nodes in the triangle are maximal, i.e.
\begin{equation*}
 \cor[(a)]{0,\theta}\prn{1,2} = \cor[(a)]{0,\theta}\prn{2,3}= \cor[(a)]{0,\theta}\prn{3,1}= \abs{\evec{\theta,0}(1)}^2 .
\end{equation*}
Hence, this triangle will be considered as a flux community.

In the case of an eigenspace of dimension larger than one, i.e. when $\eval{\ell}$ is degenerate, the relevant matrix element is $\projl_{i,j} \Tr{i}{j}$, where $\projlh$ is the projector on this eigenspace. Hence, we define in this case
\begin{equation*}
 \adj{\theta,\ell}\prn{i,j} = \abs{\projl_{i,j}} + \real{\projl_{i,j}\Tr{i}{j}} ,
\end{equation*}
for all $i,j \in V$ such that $\set{i,j} \in E$ and $\adj{\ell,\theta}\prn{i,j} = 0$ otherwise.
In practice, eigenspaces are rarely exactly degenerate. However, some eigenvalues may be approximately equal if a threshold is defined. The same issue arises in the case of spectral clustering of the combinatorial Laplacian. In order to circumvent this difficulty, it is possible to avoid the computation of the eigenvalues and use a method based on a stability criterion with respect to a change in a parameter as in~\cite{MarkovStabilityPNAS}. A similar idea can be used here, as explained in Section~\ref{SecMultiScale}.

\begin{figure}[t]
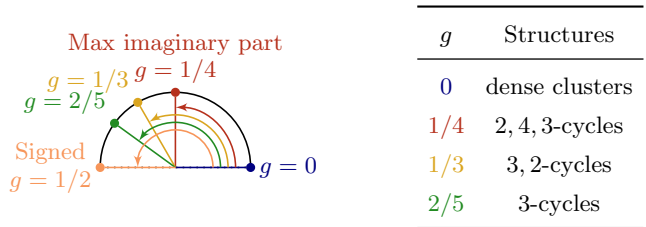

\subfloat{\adjustbox{valign=m}{\includetikznb{SchemeGValue}}}
\hfill%
\subfloat{
\begin{tabular}{cc}
 \toprule
 $g$ & Structures\\
 \midrule
 \color{graphic1s} 0 & dense clusters\\
 \color{graphic2s} 1/4 & $2,4,3$-cycles\\
 \color{graphic3s} 1/3 & $3,2$-cycles\\
 \color{graphic4s} 2/5 & $3$-cycles\\
 \bottomrule
\end{tabular}
}
\caption{Different particular values of the electric charge $\theta = 2 \pi g$. The lower half circle correspond to a network with opposite link directions. \label{PlotGValues}}
\end{figure}

Let us explain now why choosing special values of $\theta$ may be interesting. We shall focus on the first eigenvector with the smallest eigenvalue, the minimizer of~\eqref{VariationalPrinciple}.
Considering again the example of the directed triangle illustrated in Figure~\ref{FigParallelTransportTriangle} and combining the three equations of \ref{EqTriangleEquations}, we find the condition of flux quantisation
\begin{equation*}
 \Tr{1}{2} \Tr{2}{3} \Tr{3}{1} = 1 ,
\end{equation*}
i.e. $\exp\prn{\im \theta \Phi\prn{1,2,3}} = 1$ with $\Phi\prn{1,2,3} = a\prn{1,2} + a\prn{2,3} + a\prn{3,1}$.
There is a similar relation for directed $n$-cycles, which is called ``consistency'' in the case of connection graphs in~\cite{ChungZhao}. Therefore, if the links are in only one direction, i.e. $a\prn{i,j} = \pm 1$, and not reciprocal, then $\theta$ is taken such that $\theta \Phi\prn{1,2,3, \dotsc, n} = 0 \mod{2 \pi}$. This means that we can choose the parameter $\theta$ to take quantised values,
\begin{equation*}
 \theta = \frac{2 \pi k}{n}, \quad \frac{k}{n} \notin \Z ,
\end{equation*}
in order to detect $n$-cycles with constant phase differences.
A trivial consequence is that the value $\adj{\theta,\ell}\prn{i,j}$, on an edge which is not part of a flux community, will be suppressed. Hence, the unitary factor in~\eqref{Bilinear} implementing the parallel transport on the line bundle is fundamentally important. In quantum mechanics, for instance in the case of Abrikosov vortices in Type II superconductors, it is well-known that the magnetic flux is quantised. We have here an analogous condition on the product of the ``electric'' charge and the magnetic flux. Moreover, in order to detect communities of the size given by a multiple of $n$, i.e. communities of a given magnetic flux, we prescribe to choose the quantised charge $\theta = 2 \pi / n$, as shown in Figure~\ref{PlotGValues}. In the sequel, we will define the coupling constant 
\begin{equation*}
 g = \frac{\theta}{2\pi} .
\end{equation*}

\subsection{Normalized magnetic Laplacian}

The magnetic Laplacian was originally constructed in the case of a quantum particle on a lattice where, of course, the degree of the nodes is constant. Real-life networks have often an inhomogeneous degree distribution. It may be interesting to use instead of the magnetic Laplacian of~\eqref{MagneticLaplacian} a degree normalized version
\begin{equation}
\label{NormalizedMagneticLaplacian}
 \laphn{a,\im \theta} = \degs^{-1/2} \circ \laph{a,\im \theta} \circ \degs^{-1/2} ,
\end{equation}
with the degree $\degs\prn{i} = \sum_j \ws\prn{i,j}$ associated to the symmetrized weight.
The normalization can be understood as changing the definition of the covariant derivative by $\cder{a,\theta} \to \cder{a,\theta}^\text{N}  = \cder{a,\theta} \circ \degs^{-1/2}$, which can be alternatively understood as a change of measure~\cite{Blanchard} or reweighting the inner product $\inp{\psi | \psi'}_{0,N} \triangleq \sum_{i \in V} \degs\prn{i} \psi^*\prn{i} \psi'\prn{i}$, so that nodes with a large degree have a larger weight. Loosely speaking, the upshot is that nodes with a large degree are effectively considered as a set of several nodes with a smaller degree. Notice that this construction is similar to the normalized version of the combinatorial Laplacian.

\section{Clustering the auto-correlation of the magnetic eigenmaps}
\label{SecClustering}

Our proposal can be summarized as follows: from the directed  network, a weighted undirected graph is constructed, so that the weights of the links emphasize certain structures describing flux communities. This novel weighted network can then be studied with a community detection method. In this paper, we propose the use of undirected modularity, however other methods could as well be used, with possibly different outcomes.

\subsection{\texorpdfstring{Directed networks and $g \neq 0$}{Directed networks and g!=0}}

Fixing a partition of the network $C$ constituted of communities $c \in C$, we propose to cluster the network by maximizing
\begin{equation*}
 \parti{\theta,\ell}\prn{C} = \sum_{c \in C} \sum_{i,j \in c} \prn{\adj{\theta,\ell}\prn{i,j}-p_{\theta,\ell}\prn{i,j}} ,
\end{equation*}
with the configuration null model $p_{\theta,\ell}\prn{i,j} = k_i k_j / 2 m$, with $k_i = \sum_j \adj{\theta,\ell}\prn{i,j}$ and $2 m = \sum_i k_i$.
As a consequence, it is possible to use a generalized Louvain method~\cite{BlondelLouvain} to find the optimal partition based on the following customary formulation in terms of matrices:
\begin{equation}
\label{Quality}
 \parti{\theta,\ell}\prn{C} = \tr{\mH^T \brq{\mathbf{X}_{\theta,\ell} - \mathbf{p}_{\theta,\ell}} \mH} ,
\end{equation}
where $\mH\in \R^{N\times \abs{C}}$ is an indicator matrix of the communities in the partition, whose element $H_{i,j} = 1$ if node $i$ belongs to community $j$ and $H_{i,j} = 0$ otherwise.
Finding the partition maximizing the quality function~\eqref{Quality} is possible thanks to a greedy optimization algorithm\footnote{We used the code available at \url{http://netwiki.amath.unc.edu/GenLouvain/GenLouvain}), called \texttt{genLouvain}, used in~\cite{ScienceMucha}}.

\subsection{\texorpdfstring{Undirected networks and $g = 0$}{Undirected networks and g = 0}}

In the case of an undirected graph, i.e. $a = 0$, this spectral clustering method yields the condition $\pevec{\theta,\ell}\prn{i} \approx \pevec{\theta,\ell}\prn{j} \mod{2 \pi}$ for any $i$ and $j$ in the same cluster, where $\pevec{\theta,\ell} = \phase{\evec{\theta,\ell}}$.
Therefore, because the eigenvectors can be made real, this procedure is equivalent to the grouping of nodes in two communities according to $\sign{\evec{\ell}}$, which is the so-called Fiedler partition.
Moreover, in the case $a = 0$, the matrix elements in~\eqref{PositiveMatrixElements} reduces simply to
\begin{equation*}
 \adj[(0)]{0,\ell=0}\prn{i,j} \propto \delta\prn{\set{i,j} \in E} ,
\end{equation*}
for any node $i$ and $j$, so that $\adj[(0)]{0,\ell=0}$ is a binarized form of the symmetrized weight matrix $\ws$. Therefore, optimizing the modularity of the similarity matrix elements~\eqref{PositiveMatrixElements} using~\eqref{Quality} is equivalent to the modularity optimization of the undirected graph defined as the binarized skeleton of the directed network.

\subsection{\texorpdfstring{Signed Laplacian and $g=1/2$}{Signed Laplacian and g=1/2}}

\begin{figure}[t]
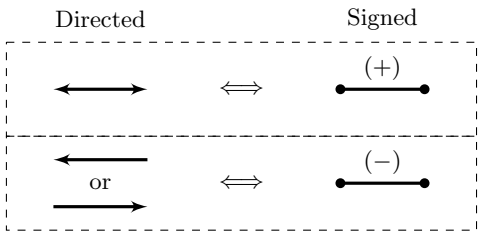

 \includetikznb{SchemeDirUndir}
 \caption{The magnetic Laplacian with $\theta = \pi$ is equivalent to a signed Laplacian of the undirected signed network obtained as illustrated above. \label{FigSignDir}}
\end{figure}

For an electric charge $\theta = \pi$, the magnetic Laplacian is actually equal to a so-called signed Laplacian~\cite{Kunegis}, associated to a specific signed network. Indeed, the parallel transporter is then real $\Tra{\pi}{j}{i} = \exp\prn{\im \pi a\prn{j,i}} = \Tra{\pi}{i}{j}$, so that $T^{\pi}_{j\to i} = 1$ if the link $\brq{i,j}$ is reciprocal, and $T^{\pi}_{j\to i} = -1$ if the link $\brq{i,j}$ is only in one direction.
In fact, the peculiar value $\theta = \pi$ can be treated as a signed network were all reciprocal links are positive whereas all other links are negative (see Figure~\ref{FigSignDir}). Therefore, our community detection method will uncover dominantly the regions of reciprocal edges, i.e. $2$-cycles.

\subsection{\texorpdfstring{The particular case $g=1/4$}{The particular case g=1/4}}

Considering for simplicity a non-degenerate eigenvalue $\eval{\ell}$, we study the particular case $g = 1/4$, so that the parallel transporter is pure imaginary $\Tra{\pi/2}{j}{i} = \im a(j,i)$. Indeed, this case corresponds to the maximal deformation of the combinatorial Laplacian in the complex domain.
The understanding of the importance of the edge directions is made easier because $a\prn{i,j} = \pm 1$ indicates the directions of $\brq{i,j}\in E$. We pretend that most of the effect of the edge directions is encoded in this difference of Aharonov-Bohm phases, that we have to compare with the sign of $a\prn{i,j}$ indicating the directionality.
Actually, the matrix elements of~\eqref{PositiveMatrixElements} satisfy
\begin{equation}
\label{CasePi/2}
 \frac{\adj{\pi/2,\ell}\prn{i,j}}{\abs{\evec{\theta,\ell}\prn{i}}\abs{\evec{\theta,\ell}\prn{j}}} = 1 + a\prn{i,j} \sin\prn{\Delta \pevec{\pi/2,\ell}\prn{i,j}} ,
\end{equation}
with the phase difference $\Delta \pevec{\pi/2,\ell}\prn{i,j} = \pevec{\pi/2,\ell}\prn{j} - \pevec{\pi/2,\ell}\prn{i}$.
Let us discuss the interest of this particular expression.
If there is a directed edge from $i$ to $j$, i.e. $a\prn{i,j} = 1$, then the value of~\eqref{CasePi/2} will be large if the phase difference is small, i.e. $0 \leq \Delta\pevec{\pi/2,\ell}\prn{i,j} < \pi$. A large value of $\adj{\pi/2,\ell}\prn{i,j}$ is expected if the nodes $i$ and $j$ belong to the same flux community. Otherwise, if the phase difference is too large, the value of~\eqref{CasePi/2} will be smaller and it will be more likely that the nodes $i$ and $j$ belong to different flux communities.
Hence, the value of the charge $\theta = \pi / 2$ is plausibly going to be the most robust choice for uncovering flux communities of different sizes. In the sequel, this value will always be chosen in the absence of reciprocal links. In the presence of reciprocal links, i.e. $2$-cycles, the choice $\theta = \pi / 2$ seems to give a large relative weight to the links satisfying $a\prn{i,j} = 0$. If, for instance, we want to give more importance to directed $3$-cycles, the choice $\theta = \pi / 2$ may not be appropriate (actually, in real-life networks directed triangles, and more generally motifs, are thought to be very important for our understanding of social or biological networks).

\subsection{\texorpdfstring{Emphasis of reciprocal links and $3$-cycles}{Emphasis of reciprocal links and 3-cycles}}

Recently, community detection techniques emphasizing the important of specific local structures have been proposed. For instance, it has been suggested to improve the methods relying on the density of links by incorporating information about triangles in the network~\cite{SerrourArenasGomez,Prat-Perez:2012}. Furthermore, directed triangles in directed graphs were shown to be relevant for community detection in~\cite{Benson-2015-tensor,Klymko-2014-dircomm}.

The edge flow $1$-form was defined so that $a\prn{i,j} = \pm 1$ on a directed link, and $a\prn{i,j} = 0$ on a reciprocal link. Let us denote the electric charge $\theta = 2\pi g$. It is clear that only the product $2 \pi g a\prn{i,j}$ is important, and we have shown that, for efficiency, the value of the charge has to be a ratio of integers. Choosing $g > 1/2$ is equivalent to a flip of the directions of the directed links, and therefore, we restrict ourselves to $g < 1/2$.  
Actually, the structures such as directed $n$-cycles will be emphasized by a choice  $g = k/n$. In practice, we assume that directed $n$-cycles with $n \leq 5$ constitute the most significant structures in the context of community detection. Incidentally, directed triangles seem to incorporate a lot of information as explained in the recent papers \cite{Benson-2015-tensor,Klymko-2014-dircomm}.

\begin{figure}[t]
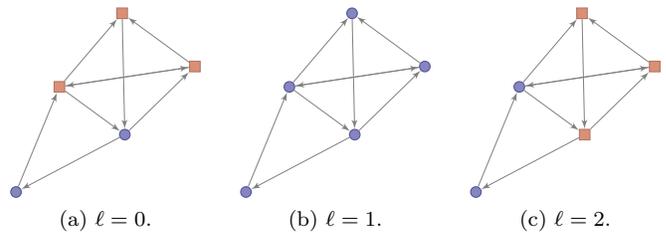

 \tikzwidth{0.33\columnwidth}
 \subfloat[$\ell = 0$.]{\plotgraph\includetikz{DoubleArrowg025ell0}}\hfill%
 \subfloat[$\ell = 1$.]{\plotgraph\includetikz{DoubleArrowg025ell1}}\hfill%
 \subfloat[$\ell = 2$.]{\plotgraph\includetikz{DoubleArrowg025ell2}}
 \caption{Partitions of an artificial network with a reciprocal link for $g= 1/4$, with the normalized magnetic Laplacian.\label{DoubleArrow1/4}}
\end{figure}

The choice that we recommend for networks without reciprocal links is $g = 1/4$, because it corresponds to the deformation of the combinatorial Laplacian with maximal imaginary part. We observe empirically that, if reciprocal links are present, the value $g = 1/4$ will give a relatively large weight to reciprocal links, since they are associated to directed $2$-cycles, as illustrated in Figure~\ref{DoubleArrow1/4}, where the reciprocal link is more emphasized that the two directed triangles of the network.

\begin{figure}[t]
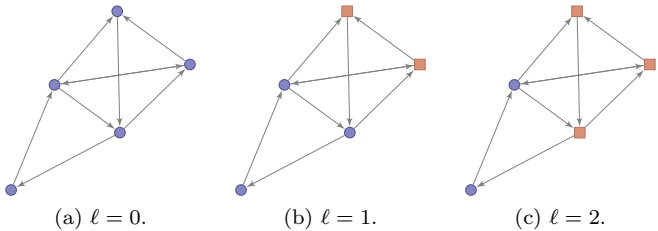

 \tikzwidth{0.33\columnwidth}
 \subfloat[$\ell = 0$.]{\plotgraph\includetikz{DoubleArrowg04ell0}}\hfill%
 \subfloat[$\ell = 1$.]{\plotgraph\includetikz{DoubleArrowg04ell1}}\hfill%
 \subfloat[$\ell = 2$.]{\plotgraph\includetikz{DoubleArrowg04ell2}}
 \caption{Partitions of an artificial network with a reciprocal link for $g= 2/5$, with the normalized magnetic Laplacian.\label{DoubleArrow2/5}}
\end{figure}

To avoid the emphasis of reciprocal links, the solution may be to choose $g = k/n < 1/2$ with $k \geq 1$, so that the edge flow on the directed links is effectively rescaled by the positive constant $k$. Heuristically, the result is that
$a\prn{i,j} \to \tilde{a}\prn{i,j} = k a\prn{i,j}$ on a directed link, and $a\prn{i,j} = 0$ on a reciprocal link. Restricting ourselves to $n \leq 5$ and $g < 1/2$, we then recommend to choose $g = 2/5$. The results of this choice of an artificial network is displayed in Figure \ref{DoubleArrow2/5}, where the directed triangles are more emphasized than the reciprocal link.
The structures found with different $g$  values are summarized in Figure \ref{PlotGValues}.

\begin{figure}[t]
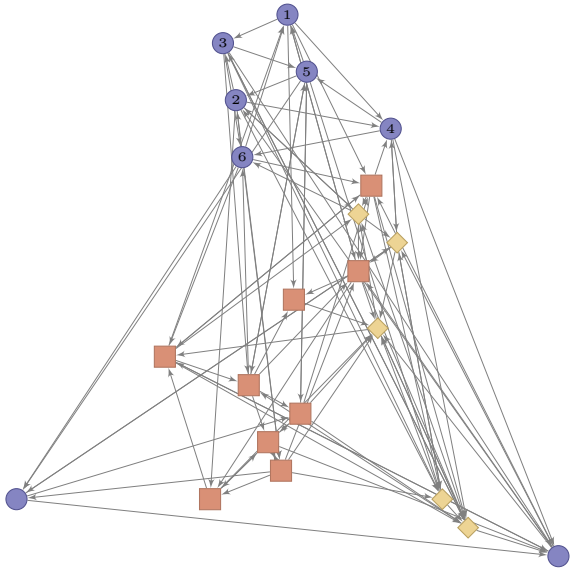

 \tikzwidth{\columnwidth}
 \plotgraph\includetikz{Anomalous}
 \caption[]{Example of a directed network with a planted community~\cite{Benson-2015-tensor} and reciprocal edges. The numbered nodes (1--6) are part of many directed triangles. The rest of the nodes follow an Erd\`os-R\'enyi pattern. Our method ($\ell = 0$ and $g = 2/5$) detects this planted community and includes two other nodes which are also part of directed cycles. The same ``anomalous'' community is detected for $\ell = 0$ and $g = 1/3$. \label{BensonAnomalous}}
\end{figure}

Another network with a planted structure and including reciprocal links was proposed in~\cite{Benson-2015-tensor}\footnote{Available at \url{https://github.com/arbenson/tensor-sc}.}. Our method with $g = 2/5$ is able to uncover the so-called ``anomalous'' community constituted of a set of $6$ nodes connected by many directed triangles while the rest of the network is randomly generated, as illustrated in Figure~\ref{BensonAnomalous}.

\section{Flux communities at finite temperature}
\label{SecMultiScale}

The methods of the previous section rely on the computation of the spectrum of the magnetic Laplacian. We may expect that the larger is the eigenvalue, the finer is the structure detected. The exploration of the multiscale structure of the network may be performed by considering the same quantum mechanical problem but at finite temperature, the latter parameter being used as a scale parameter.

\subsection{Density operator}

Let us recall the basics of quantum mechanics at finite temperature~\cite{Wipf}.
Consider a directed network and an Hamiltonian operator on the complex valued functions of the nodes. The Hamiltonian $\Hh$ defines a mixed state representing a statistical distribution of excitations at a certain temperature $T = 1 / \beta$, given by the so-called density operator (or density matrix), i.e.
\begin{equation}
\label{Density_Matrix}
 \dmath{\beta} = \frac{e^{ -\beta \Hh}}{Z\prn{\beta}} , \quad \text{with } Z\prn{\beta} = \tr{e^{ -\beta \Hh}} .
\end{equation}
This is well-known to be analogous to quantum mechanics with imaginary time. The statistical mixture is easily understood using the spectral representation of the Hamiltonian in Dirac notation,
\begin{equation*}
 \dmath{\beta} = \sum_{\ell = 0}^{N-1} \frac{e^{-\beta \eval{\ell}}}{Z\prn{\beta}} \ket{\evec{\ell}}\bra{\evec{\ell}} ,
\end{equation*}
where the coefficient of each term in the sum is a probability representing the proportion of the eigenvector $\ket{\evec{\ell}}$ in the mixed state. For simplicity, the eigenvalues are sorted in ascending order.

\subsection{Correlation at finite temperature}

We choose the Hamiltonian to be the positive semi-definite magnetic Laplacian $\Hh = \laph{a,\im \theta}$.
Let us define the matrix elements $\dmat{\beta}{i,j} = \inp{ \delta_i |\dmath{\beta} |\delta_j }$. Incidentally, the matrix elements of the density operator are proportional to the Euclidean time Feynman propagator $K\prn{j\to i,\beta} = \inp{ \delta_i | e^{- \beta \Hh}|\delta_j}$, between the node $j$ to $i$ (for a reference, see~\cite{Wipf}).

The connection with the approach of the previous section is more clear when writing
\begin{equation*}
 \Tr{i}{j} \dmat{\beta}{i,j} = \sum_{\ell = 0}^{N-1} \frac{e^{- \beta \eval{\ell}}}{Z\prn{\beta}} \prn{\evec{\theta,\ell}\prn{i} \Tr{i}{j} \evect{\theta,\ell}\prn{j}} ,
\end{equation*}
which is a weighted sum of the correlations appearing in~\eqref{CorrelationTransport} for each eigenspace.
Therefore we introduce the positive matrix elements at inverse temperature $\beta$,
\begin{equation*}
 \adj{\beta,\theta}\prn{i,j} = \abs{\dmat{\beta}{i,j}} + \real{\Tr{i}{j} \dmat{\beta}{i,j}} .
\end{equation*}
In the low temperature limit, we obtain the correlation~\eqref{PositiveMatrixElements} corresponding to the lowest energy state $\ell = 0$, i.e.
\begin{equation*}
\lim_{\beta \to \infty} \adj{\beta,\theta}\prn{i,j} = \adj{\ell = 0,\theta}\prn{i,j} .
\end{equation*}

As a consequence, $\adj{\beta,\theta}$ can be viewed as the weighted similarity matrix of an undirected network. As before, the quality function of a partition is simply chosen to be the modularity
\begin{equation*}
 \parti{\beta,\theta}\prn{C} = \sum_{c\in C} \sum_{i,j \in c} \prn{\adj{\beta,\theta}\prn{i,j} - p_{\beta,\theta}\prn{i,j}} ,
\end{equation*}
with the null model $p_{\beta,\theta}\prn{i,j}$ chosen to be the configuration model.

\section{Flux communities in artificial and real-life networks}
\label{SecFluxARNet}

\begin{figure*}[t]
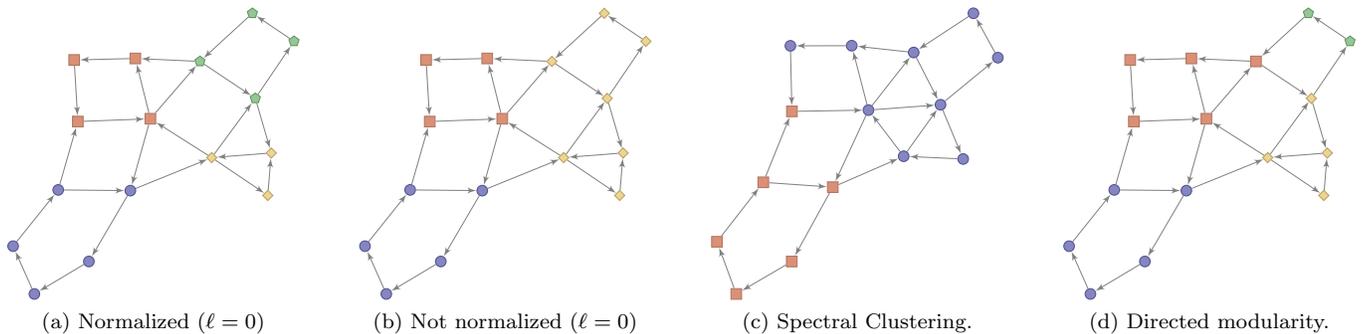

 \tikzwidth{0.25\textwidth}
 \subfloat[Normalized ($\ell = 0$)]{\plotgraph\includetikz{VariousSizeNorm}\label{VariousSizeNorm}}\hfill%
 \subfloat[Not normalized ($\ell = 0$)]{\plotgraph\includetikz{VariousSizeNotNorm}\label{VariousSizeNotNorm}}\hfill%
 \subfloat[Spectral Clustering.]{\plotgraph\includetikz{VariousSizeSpectralClustering}\label{VariousSizeSpecClust}}\hfill%
 \subfloat[Directed modularity.]{\plotgraph\includetikz{VariousSizeDirectedModularity}\label{VariousSizeDirMod}}
 \caption{Comparison of our methods for $\ell = 0$ and $g=1/4$. The normalized magnetic Laplacian of~\eqref{NormalizedMagneticLaplacian} may allow to uncover communities in networks with an inhomogeneous degree distribution. In Figure~\ref{VariousSizeSpecClust}, we show the result of spectral clustering based on the normalized combinatorial Laplacian (Fiedler partition). Directed modularity~\cite{LeichtNewman} does not uncover the same communities, as illustrated in Figure~\ref{VariousSizeDirMod}. \label{VariousSize}}
\end{figure*}

Let us first study the effect of the normalization on an artificial example.
The results obtained using the normalized and un-normalized magnetic Laplacians are compared on an example with flux communities of different sizes, illustrated in Figure~\ref{VariousSize}, where the normalized method is able to distinguish all the communities.

\begin{figure}[t]
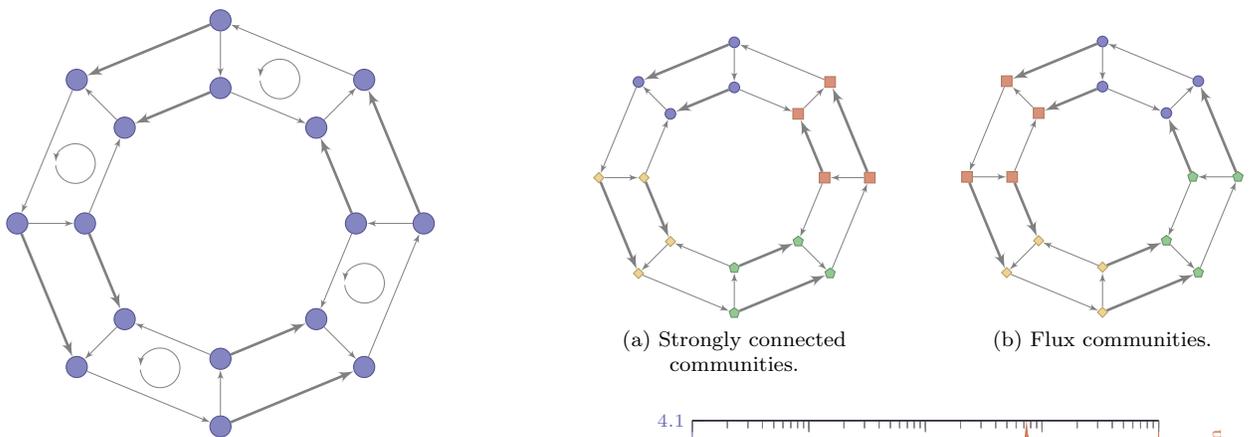

 \tikzwidth{0.75\columnwidth}
 \centering
 \plotgraph\includetikz{GroupOf4}
 \caption{A directed networks with flux communities. The network is formed by four groups of $4$ nodes forming a directed cycle with unit weights. These groups are connected by edges of weight $2$, displayed in bold. The four directed cycles with nodes of the same color form four flux communities of $4$ nodes. \label{GroupOf4}}
\end{figure}

Flow communities are often defined as structures retaining the flow of a dynamical process~\cite{MarkovStabilityIEEE}. Here, the dynamics is not given by a Markov process, so our interpretation leads us to name them ``flux'' communities. A typical example is depicted in Figure~\ref{GroupOf4}. This toy directed network has been studied using LinkRank (directed modularity)~\cite{Kim}, Markov Stability~\cite{MarkovStabilityIEEE}, and Infomap~\cite{RosvallBergstrom}, and it is constituted of four flux communities, which are directed cycles. The difficulty to detect them is that the edges interconnecting the groups have a double weight, so that a clustering based on the symmetrized weight matrix will find four different communities of nodes connected by those strong links. In~\cite{Kim}, this network is studied in order to describe an example where the directed modularity definition of~\cite{LeichtNewman} is unable to discover the communities. These structures are however highlighted by the information theory framework of~\cite{RosvallBergstrom}, whereas the Markov Stability framework is able to uncover them using a random walk with teleportation as a means to explore the network.

\begin{figure}[t]
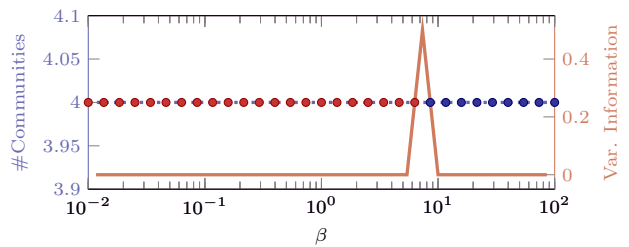

 \tikzwidth{0.5\columnwidth}
 \subfloat[Strongly connected communities.]{\plotgraph\includetikz{GroupOf4StrongGroups}\label{GroupOf4NextExcited}}\hfill%
 \subfloat[Flux communities.]{\plotgraph\includetikz{GroupOf4FirstExcited}\label{GroupOf4FirstExcited}}\\
 \centering
 \tikzwidth{0.9\columnwidth}
 \subfloat[Stability of the partitions.]{\includetikz{StabilityArt4}}
 \caption[Communities uncovered.]{The communities uncovered by our method for $g=1/4$ and using the normalized magnetic Laplacian~\eqref{MagneticLaplacian}. At low $\beta$, the $4$ strongly connected groups form the stable partition of Figure~\ref{GroupOf4NextExcited} marked as~\showscatter{1000}, while at large $\beta$, i.e. low temperature, the method uncovers $4$ flux communities ($4$-cycles in Figure~\ref{GroupOf4FirstExcited}), which is denoted by~\showscatter{0}. The variation of information is shown in order to detect the change of partition. \label{FluxCommunitiesGroupOf4}}
\end{figure}

In the framework proposed in this paper, communities are associated to a certain flux and this flux is related to a certain structure of the network. The variation of the flux is controlled by the quantised coupling constant and allows to uncover communities of different types.
Let us use the finite temperature method on the network of Figure~\ref{GroupOf4}.
Using this method for $g = 1/4$, we uncover two types of communities for the low energy states depicted in Figure~\ref{FluxCommunitiesGroupOf4}.  Since the weights are not equal in this network, the normalized Magnetic Laplacian~\eqref{NormalizedMagneticLaplacian} is used.
First of all, at high temperature, we obtain four other groups of $4$ nodes, which are more connected (see Figure~\ref{GroupOf4NextExcited}). This partition is easily uncovered in the absence of edge directions. Secondly, at a lower temperature, we detect four communities which are the $4$-cycles illustrated in~Figure \ref{GroupOf4FirstExcited}. They constitute flux communities, and they are also obtained in the references~\cite{Kim,RosvallBergstrom,MarkovStabilityIEEE}.
The robustness of the finite temperature method under a slight perturbation of the network is studied in Appendix~\ref{SecRobustness}.

Actually, directed modularity optimization discovers only the partition of Figure~\ref{GroupOf4NextExcited}. Infomap~\cite{RosvallBergstrom}, based on a random walk, is able to uncover the partition of Figure~\ref{GroupOf4FirstExcited}, and the Markov stablility framework~\cite{MarkovStabilityIEEE} finds it as well for a certain Markov time scale.

\begin{figure}[t]
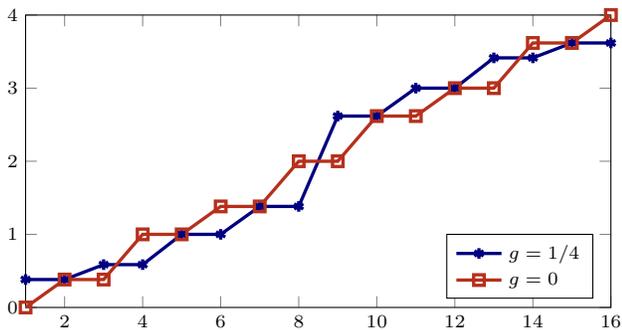

 \tikzwidth{0.9\columnwidth}
 \centering
 \includetikz{SpectrumGroupOf4}
 \caption{Spectrum of the magnetic and combinatorial Laplacian for the network of Figure \ref{GroupOf4}. Clearly, in the magnetic Laplacian the gap between the ground state and the first excited state is small, with respect to the gap between the two excited levels. \label{SpectrumGroupOf4}}
\end{figure}

In order to study the same network level by level, the eigenvalues of the magnetic Laplacian are needed. The spectrum of the magnetic Laplacian provides an indication about the quality of the communities obtained. If the gap between the lowest energy level and the first excited levels is small, the significance of the community found at the first excited level is expected to be high. This is analogous to the well-known interpretation of spectral clustering using the combinatorial Laplacian. Incidentally, we can observe that the spectrum of the combinatorial Laplacian in Figure~\ref{SpectrumGroupOf4}, built using $\ws$ as affinity matrix, is qualitatively different from the spectrum of the magnetic Laplacian.
Noticeably, in the case of a connected network, there is no information about community structure encoded in the eigenvector of minimal eigenvalue of the combinatorial Laplacian. The same is not true for the magnetic Laplacian.
In particular, the smallest eigenvalue of the magnetic Laplacian does not vanish in general and can be degenerate, i.e. the eigenspace can be of dimension greater than one. This degeneracy is intuitively very natural since in quantum mechanics in the continuum the so-called Landau levels are infinitely degenerate.

\subsection{Comparison with directed modularity}

We can compare the results given by the maximization of directed modularity with our method based on the normalized magnetic Laplacian on the two previous artificial networks. The first example in Figure \ref{FluxCommunitiesGroupOf4} shows how our method is able to detect communities allowing a magnetic flux (Figure~\ref{GroupOf4FirstExcited}) whereas directed modularity relies more on relative edge density, obtaining the partition of Figure~\ref{GroupOf4NextExcited}. In the second example of Figure~\ref{VariousSize}, our method discovers all flux communities (Figure~\ref{VariousSizeNorm}), while directed modularity gives a different partition (Figure~\ref{VariousSizeDirMod}).

\subsection{A real-file example}

\begin{figure*}[t]
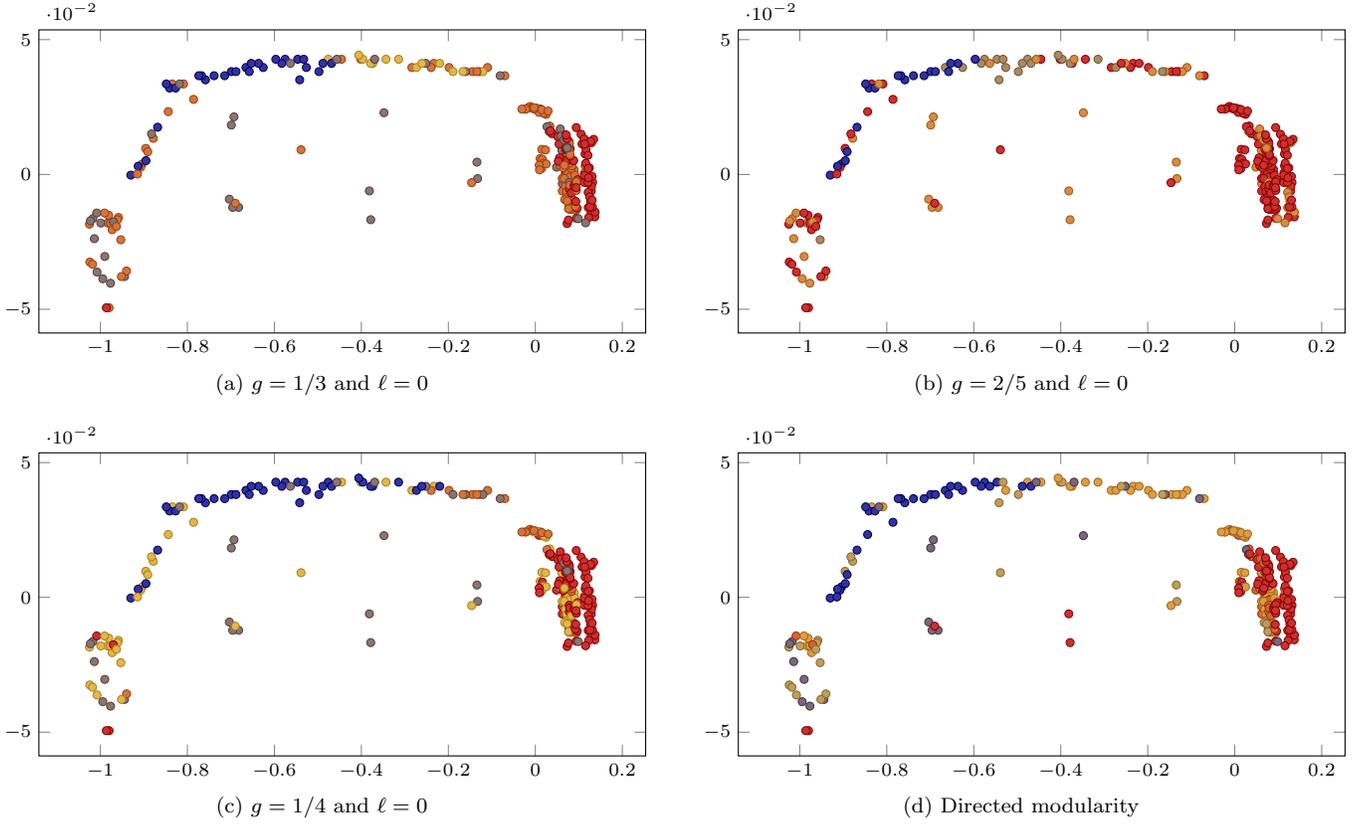

 \tikzwidth{0.45\textwidth}
 \subfloat[$g=1/3$ and $\ell =0$]{\includetikz{CElegansPositions03}}\hfill%
 \subfloat[$g=2/5$ and $\ell =0$]{\includetikz{CElegansPositions04}}\\
 \subfloat[$g=1/4$ and $\ell =0$]{\includetikz{CElegansPositions025}}\hfill%
 \subfloat[Directed modularity]{\includetikz{CElegansPositionsDirMod}}
 \caption{Communities in the C. Elegans directed network found by optimizing~\eqref{Quality}. The nodes are displayed according to the physical positions of the neurons, units are cm. In each figure, the nodes in the same community share the same color. \label{CElegans}}
\end{figure*}

To illustrate the result of our method, we study the neuronal network of the C. Elegans nematode~\cite{KaiserPLOS,KaiserHilgetagPLOS}\footnote{Data obtained from \url{http://www.dynamic-connectome.org/}.}, constituted of $277$ neurons. In Figure~\ref{CElegans}, the communities found for different values of the electric charge $g$ are visualized, using the physical spatial coordinates as positions of the neurons.
Qualitatively, the partitions found in Figure~\ref{CElegans} for the lowest energy level are similar to the partition found by optimizing the directed modularity. However, one may observe that our method seems to group into the same community the neurons appearing in the middle of the figure ( i.e. AVM, ALMR, ALML, BDUR, BDUL, PDER, PDEL, PVDR, PVDL, PVM), whereas directed modularity separates them in different communities.

\section{Spectral clustering in the complex plane and flow communities}
\label{SecClusteringComplex}

In the previous section, the importance of directed cycles for detecting dense communities was emphasized. There are however many networks whose structure resides more in the flow of link directions. The magnetic eigenmaps can also reveal such features.

Let us consider a directed network such that its edge flow $a = a_h$,
\begin{equation}
\label{ExactEdgeFlow}
 a_h = \der h ,
\end{equation}
is a pure gradient, i.e. an exact $1$-form satisfying $a_h\prn{i,j} = h\prn{j} - h\prn{i}$ for any link $\set{i,j}$. Incidentally, in the language of combinatorial Hodge theory~\cite{YaoRanking}, $a = a_h$ is a consistent ranking. Based on the gauge covariance property~\eqref{GaugeTransformation}, it is straightforward to prove that the eigenvalues and eigenvectors of the magnetic Laplacian $\laph{\der h,\im \theta}$ are in one to one correspondence with the eigenvalues and eigenvectors of the combinatorial Laplacian, i.e.
\begin{equation}
\label{MagneticPureDerivative}
 \laph{\der h,\im \theta} = e^{-\im\theta h} \circ \laphc \circ e^{\im \theta h} .
\end{equation}
Actually, the function $h$, defined on the nodes, is a potential whose gradient gives the edge flow. In particular, in the case of a connected directed network, the lowest energy state of the magnetic Laplacian~\eqref{MagneticPureDerivative} is simply given by
\begin{equation*}
\label{LowestEnergyState}
 \evec{\theta,0}\prn{i} = e^{- \im \theta h\prn{i}} \evec{0,0}\prn{i} , \quad \text{with } \evec{0,0}\prn{i} = \cst ,
\end{equation*}
where $\evec{0,0}$ is the constant eigenvector of the combinatorial Laplacian.
In this case, we propose to assign two nodes $i$ and $j$ in the same community if
\begin{equation*}
 \phase{\evec{\theta,0}\prn{i}} \approx \phase{\evec{\theta,0}\prn{j}} ,
\end{equation*}
which corresponds to a spectral clustering in the complex plane of the eigenfunction of lowest energy. Hence, the communities are the groups of nodes with the same potential $h$.

\subsection{Communities with a running flow}

In real-life networks, the edge flow rarely satisfies the exactness condition \eqref{ExactEdgeFlow}, and therefore, the eigenvectors of $\laph{a,\im \theta}$ and $\laphc$ are not exactly in one to one correspondence. Nevertheless, it is still instructive to study the lowest energy state of $\laph{a,\im \theta}$, i.e. the first magnetic eigenmap.

\begin{figure}[t]
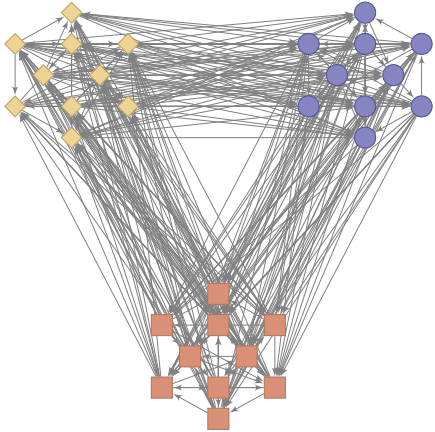

 \tikzwidth{0.75\columnwidth}
 \centering
 \plotgraph\includetikz{3Groups}
 \caption{Directed network with a flow running between three sets of nodes. The nodes within each group are not strongly connected. \label{3Groups}}
\end{figure}

An example network where the lowest eigenvector of magnetic Laplacian is able to uncover a community structure is given in Figure~\ref{3Groups}. It is composed of three groups of ten nodes. Within each group there is a uniform probability $0.5$ that two nodes are connected. There is also a probability $0.5$ that a node connects with a node from another group. However, $90$ percent of the connections between the groups point in the direction of the flow. A similar network was actually proposed in~\cite{Lancichinetti}. The communities can be detected thanks to mixture models~\cite{NewmanMixture}, whereas directed modularity is expected to fail.
Actually, this type of community structure is associated to the role played by each node in the network, and hence this feature is in the same spirit than the so-called Role Based Similarity~\cite{CooperBarahonaRBS,TwitterInterestCommunities}.

\begin{figure}[t]
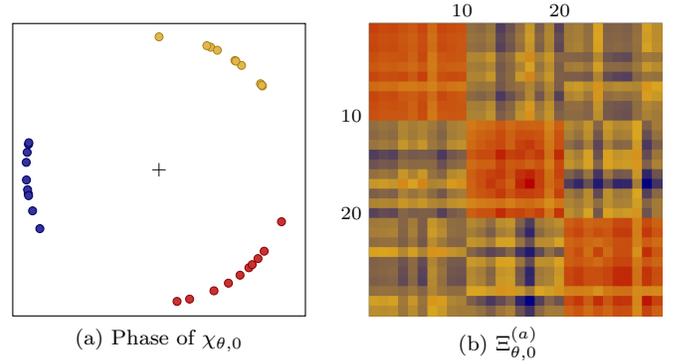

 \tikzwidth{0.45\columnwidth}
 \subfloat[Phase of $\evec{\theta,0}$]{\includetikz{FirstEigenV3Groups}}\hfill%
 \subfloat[$\Xi^{(a)}_{\theta,0}$]{\includetikz{CorrFlow3Groups}}
 \caption{Phase of the lowest eigenvector $\evec{\theta,0}$ and the matrix $\Xi^{(a)}_{\theta,0}$ for $g = 1/4$, as a function of the vertex number for the network of Figure~\ref{3Groups}. \label{FirstEigenV3Groups}}
\end{figure}

The phase of the lowest eigenvector is depicted in Figure~\ref{FirstEigenV3Groups}. Indeed, the phase of $\evec{\theta,0}$ is almost piecewise constant.

A similar situation happens for the word adjacency directed network of Figure~\ref{WordNetwork}, which was constructed from an English text~\cite{NewmanPREEigenvectors}\footnote{Data obtained from \url{http://www-personal.umich.edu/~mejn/netdata/}.}. The network is built by collecting adjacent nouns and adjectives in the novel \emph{David Copperfield}. Hence, a directed link points from a word to another adjacent word, if the first one appears before the second. From the structure of English, we can expect a certain flow structure in the network. Indeed, the phase of the first eigenmaps separates the nouns from the adjectives.

\begin{figure}[t]
 \tikzwidth{0.75\columnwidth}
 \centering
 \includetikz{WordPhasesLabel}
 \caption[Phase for word adjacency.]{The phase of the first magnetic eigenmaps for the word adjacency network and $g=2/5$, clearly separating the nouns (denoted as \showscatter{1000}) from the adjectives (denoted as \showscatter{0}), represented on a circle. \label{WordNetwork}}
\end{figure}

The difficulty of discovering these type of communities using the phase of the first eigenvector of the magnetic Laplacian is that the number of communities has to be guessed. In order to circumvent this problem, we propose the definition of a quality function. Actually, the correlations
\begin{equation*}
\label{BilinearFlow}
 \cornt[(a)]{\theta,0}\prn{i,j} = \real{\evec{\theta,0}\prn{i} \evect{\theta,0}\prn{j}} ,
\end{equation*}
for any nodes $i$ and $j$, incorporate the information necessary to find the flow communities in the network. Hence, we define the positive matrix elements
\begin{equation}
\label{PositiveMatrixElementsFlow}
 \adjnt[(a)]{\theta,0}\prn{i,j} = \abs{\evec{\theta,0}\prn{i} \evect{\theta,0}\prn{j}} + \cornt[(a)]{\theta,0}\prn{i,j} ,
\end{equation}
which define another weighted similarity matrix. The matrix $\adjnt[(a)]{\theta,0}$ corresponds to another network containing the same nodes but with new weighted undirected links. The correlation $\adjnt[(a)]{\theta,0}$ differs from $\adj[(a)]{\theta,0}$ only by the parallel transport factor.
Noticeably, this matrix is not invariant under a gauge transformation $a \to a+ \der h$.

\subsection{Optimization of a quality function}

A modularity optimization procedure on $\adjnt[(a)]{\theta,0}$ allows to uncover the three flow communities of Figure~\ref{3Groups}, depicted with different colours. For simplicity, we used the Newman-Girvan modularity~\cite{NewmanGirvan} with the configuration model, although actually we could have chosen another method in order to partition the undirected network associated to $\adjnt[(a)]{\theta,0}$.

To compare this method with directed modularity on another network, we examine now a real-life example used in~\cite{LeichtNewman} representing the Big Ten football network. In the directed network of Figure~\ref{Football}, our method based on the optimization the modularity of~\eqref{PositiveMatrixElementsFlow} gives three communities. However, directed modularity gives only two communities. In our method, there are two communities of teams shared with the directed modularity approach, however the team of Minnesota (number $4$) is singled out because its number of successes equals its number of defeats. Therefore, the partition found by our method seems to be consistent with the intuition.

\begin{figure}[t]
 \tikzwidth{0.5\columnwidth}
 \subfloat[$g=1/4$ and $\ell =0$.]{\plotgraph\includetikz{FootballNewman3Comm025}}\hfill%
 \subfloat[Directed modularity.]{\plotgraph\includetikz{FootballNewman3CommDirMod}}\\
 \centering
 \subfloat[List of the teams and their associated position.]{
 \begin{tabular}{lc|lc}
  \toprule
  Team & Position & Team & Position\\
  \midrule
   Penn State & 1 & Wisconsin & 7 \\
   NorthWestern & 2 & Illinois & 8 \\
   Ohio State & 3 & MichiganState & 9 \\
   Minnesota & 4 & Purdue & 10 \\
   Michigan & 5 & Indiana & 11 \\
   Iowa & 6 & \\
   \bottomrule
 \end{tabular}}
 \caption{Comparison of our method based on the modularity optimization of~\eqref{PositiveMatrixElementsFlow} for flow communities of the un-normalized magnetic Laplacian with directed modularity in the Big Ten football network, and the ranking of the teams. \label{Football}}
\end{figure}

\section{Conclusions}
\label{SecConlusions}

Link directions in complex networks may contain relevant information. Accounting for this information may be done in various manners, for example several dynamical processes may be imagined to explore the geometry of the networks.
In the past, much effort was devoted to the study of the geometrical structure of networks in terms of density ``clusters''. On the contrary, we have been interested here in other local structures which are also related to the topology of the networks, where the word ``topology'' is understood in the mathematical sense.

In particular, the use of the complex valued magnetic Laplacian for the problem of community detection in directed network was studied in this paper. The method was strongly inspired by quantum physics, and it generalizes known results to the complex domain.
A striking feature of the magnetic Laplacian is that it is related to the topology of the network. Indeed, there is a strong relationship between discrete Hodge theory and the results presented in this paper.
As we have illustrated with several experiments, this approach allows to unveil communities on directed graphs based either on cycles (flux communities) or in the role of the different nodes.

It is expected that different deformations of the combinatorial Laplacian may be relevant to answer other questions of interest for the study of complex networks.

\begin{acknowledgments}
 The authors would like to thank the following organizations.
 \begin{itemize*}
  \item EU: The research leading to these results has received funding from the European Research Council under the European Union's Seventh Framework Programme (FP7/2007-2013) / ERC AdG A-DATADRIVE-B (290923).  This paper reflects only the authors' views and the Union is not liable for any use that may be made of the contained information.
  \item Research Council KUL:  CoE PFV/10/002 (OPTEC), BIL12/11T; PhD/Postdoc grants.
  \item Flemish Government:
  \begin{itemize*}
   \item FWO: projects: G.0377.12 (Structured systems), G.088114N (Tensor based data similarity); PhD/Postdoc grant.
   \item iMinds Medical Information Technologies SBO 2015.
   \item IWT: POM II  SBO 100031.
  \end{itemize*}
  \item Belgian Federal Science Policy Office: IUAP P7/19 (DYSCO, Dynamical systems, control and optimization, 2012-2017).
 \end{itemize*}
\end{acknowledgments}

\bibliography{References}
\bibliographystyle{unsrt}

\appendix

\section{Relation with previous work}

\subsection{Discrete Vector Bundle Laplacian}
\label{SecDVBL}

From a gauge theory perspective, the each of the factors $\exp\prn{\im \theta a\prn{i,j} / 2}$ may be understood intuitively as a unitary ``parallel transport'' along the edge from $i$ to the mid-point between $i$ and $j$. A parallel transport is an isomorphism between the fibres of a vector bundle. Hence, the magnetic Laplacian~\eqref{MagneticLaplacian} has a covariance property~\cite{deVerdiere} under $a \to a + \der  h$. Let us outline this idea.

In the reference~\cite{KenyonVectorBundle}, Kenyon defines a vector bundle $V_G$ on a graph as the choice of a vector space $V_v$, called fibre, for each vertex $v \in V$. Here, for simplicity we choose $V_v$ to be isomorphic to $\C$. A section of the vertex bundle is in fact given by one complex number for each vertex, hence it is a $0$-form in $\Omega_{0}$.

Furthermore, Kenyon extends the construction to the edge space, so that there is a fibre isomorphic to $\C$, for each edge. If the edges are oriented, a $1$-form is a section of the edge bundle, i.e. a skew symmetric function of $\Omega_{1}$ in our notations. Then, it is possible to define a connection isomorphism $\phi_{ve} = \phi_{ev}^{-1}$ for each edge $e = [v,v']$ and vertex at the boundary of the edge. In our case, we choose
\begin{equation*}
 \phi_{je} = e^{\im\theta a\prn{i,j}/2}, \quad \text{for } e = \brq{i,j} .
\end{equation*}
Still, following~\cite{KenyonVectorBundle}, we introduce the map $\der : \Omega_{0} \to \Omega_{1}$
\begin{equation*}
 \prn{\der \psi}\prn{e} = \phi_{je} \psi\prn{j} - \phi_{je}\psi\prn{i} .
\end{equation*}
Subsequently, the operator $\der^\star : \Omega_{1} \to \Omega_{0}$ is introduced by
\begin{equation*}
 \prn{\der^\star \omega}\prn{i} = \sum_{j|e = \set{i,j}} \ws\prn{e} \phi_{ej} \omega(e) .
\end{equation*}
Hence, $\der^\star$ is the adjoint of $\der$ only if $\phi_{ve} = \phi_{ev}^{*} = \phi_{ev}^{-1}$.
However, because we wish a self-adjoint Laplacian,  the construction of the Laplacian of~\cite{KenyonVectorBundle} as $\der^\star \der $ is the one of this paper only if $\phi_{ev}^{*} = \phi_{ev}^{-1}$, leading to the magnetic Laplacian of~\eqref{MagneticLaplacian} and studied in~\cite{Berkolaiko,deVerdiere, Shubin}.

\subsection{Connection Laplacian}
\label{SecMLasCL}

Let us outline the relation between the discrete connection Laplacian~\cite{SingerWu} and the magnetic Laplacian.
Since $\gU{1}$ and $\gSO{2}$ are isomorphic, another definition of the magnetic Laplacian can be
\begin{equation*}
 \prn{\laph{a,\theta} \bpsi}\prn{i} = \sum_{j}\ws\prn{i,j} \prn{\bpsi\prn{i} - \brho{j \to i} \bpsi\prn{j}} ,
\end{equation*}
with the matrix of the lowest dimensional representation of $\gSO{2}$,
\begin{align*}
\brho{j\to i} &= e^{\im\theta a(j,i)\mathfrak{m}} \\
 &= \begin{pmatrix}
     \cos\theta a(j,i) & -\sin\theta a(j,i)\\
     \sin\theta a(j,i) & \hphantom{-}\cos\theta a(j,i)
    \end{pmatrix}
\end{align*}
with the skew-symmetric matrix
\begin{equation*}
 \mathfrak{m} = \begin{pmatrix}
                 0 & -1\\
                 1 & 0
                \end{pmatrix} ,
\end{equation*}
while $a\prn{i,j}$ takes the values $0$, $-1$ or $+1$.

A major difference with respect to the reference~\cite{SingerWu} is that, here, we choose the representation of $\gU{1}$ in order to detect specific structures in directed network, whereas Singer and Wu simply build from a dataset an undirected network with a matrix of $\gO{d}$ on each link. The question of choosing a representation of $\gO{d}$ is therefore not relevant.

\section{Computational aspects}

The methods presented here include two steps. The methods focusing on individual eigenvectors requires the computation of the smallest eigenvalue. Because we have noticed that the normalized magnetic Laplacian $\laphn{a,\im \theta}$ was empirically more successful, we can merely compute the maximal eigenvalues of the following operator $\hat{\mathcal{S}}_{a,\im \theta} = \iden - \laphn{a,\im \theta}$, for instance thanks to the power method.
The matrix exponential~\eqref{Density_Matrix} can be computed thanks to a Pad\'e approximant method, see for instance the scaling and squaring method~\cite{MatExp}.
For the maximization of the quality functions, we rely on the generalized Louvain method~\cite{ScienceMucha,BlondelLouvain}.

\section{Robustness of the finite temperature method}
\label{SecRobustness}

\begin{figure}
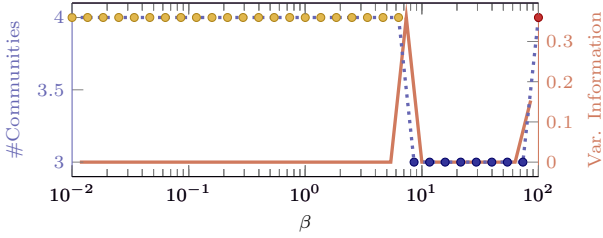

 \tikzwidth{0.33\columnwidth}
 \subfloat[Strongly connected communities.]{\plotgraph\includetikz{FirstPartCorrupted}\label{FirstPartCorrupted}}\hfill%
 \subfloat[Flux communities.]{\plotgraph\includetikz{SecondPartCorrupted}\label{SecondPartCorrupted}}\hfill%
 \subfloat[Flux communities ($T\approx 0$).]{\plotgraph\includetikz{ThirdPartCorrupted}\label{ThirdPartCorrupted}}
 \centering
 \tikzwidth{0.9\columnwidth}
 \subfloat[Stability of the partitions.]{\includetikz{StabilityArt4Corrupted}}
 \caption[Communities uncovered for the corrupt graph.]{The communities uncovered by our method for $g = 1/4$ and using the normalized magnetic Laplacian \eqref{MagneticLaplacian} (i.e. \showscatter{500} correspond to the partition of Figure~\ref{FirstPartCorrupted}, \showscatter{0} to Figure~\ref{SecondPartCorrupted}, and \showscatter{1000} to the partition of Figure~\ref{ThirdPartCorrupted}). The red directed link in the network is the flipped link compared to Figure~\ref{GroupOf4}. \label{StabilityArt4Corr}}
\end{figure}

In order to test the robustness of the partitions found in the network of Figure~\ref{GroupOf4}, we have simply slightly modified the direction of one link of weight $1$ in one of the $4$-cycles of this network, which is marked in red in Figure~\ref{StabilityArt4Corr}. Hence, one of the four $4$-cycles is broken. As a result, we find at high temperature the same partition as in the case of Figure~\ref{GroupOf4}, which is due to the presence of the links of weight $2$. In that case, the cycles are broken.
At lower temperature, the stable partition is given by groups containing the three unbroken $4$-cycles, showing that the method reacts well when one $4$-cycle is perturbed. Finally, at a very small temperature, where the importance of the flux is the largest, another partition preserving the cycles is obtained, with a fourth community found at the perturbed link.

\end{document}

%% file: Plots.tex
\newif\iftikz

\iftikz\usetikzlibrary{external,arrows,positioning,shapes,decorations.markings,arrows.meta,backgrounds}\fi

\iftikz\pgfplotsset{compat=1.13}\fi

\newcommand{\plotgraph}{\pgfplotsset{axis lines=none}}

\tikzstyle{light}=[draw,circle,graphic1,left color=graphic1,minimum width=10pt,inner sep=0pt,text=black]
\tikzstyle{lightSmall}=[light,minimum width=7pt]
\tikzstyle{lightVSmall}=[light,minimum width=5pt]
\tikzstyle{lightVSmall2}=[lightVSmall,graphic5,left color=graphic5]
\tikzstyle{linkexample}=[->,very thick, graphic1s]
\tikzstyle{arcexample}=[gray,->]
\tikzstyle{arclinkexample}=[linkexample,thick]
\tikzstyle{linkexample2}=[->,very thick, graphic5s]
\tikzstyle{shadowexample}=[fill=graphic1,fill opacity=0.15,draw opacity=0]
\tikzstyle{shadowexample2}=[shadowexample,fill=graphic5]
\tikzstyle{valuevect}=[-,very thick,graphic2s]
\tikzstyle{valuenode}=[draw,fill,circle,minimum width=0pt,inner sep=1pt,graphic2s]
\tikzstyle{centrnode}=[draw,fill,circle,minimum width=0pt,inner sep=1pt,graphic1s]

\tikzset{vertex/.style={draw=.!80!black, fill=.!80!white, inner sep=0pt, minimum width={width("9")}, minimum height={width("9")}, text = Black}}

\newcommand{\widthsm}{3pt}

\tikzset{vertexcol/.style={color of colormap={#1}, shape=circle, vertex}}

\tikzset{vertextyp1/.style={color=graphic1, shape=circle, vertex}}
\tikzset{vertextyp2/.style={color=graphic2, shape=rectangle, vertex}}
\tikzset{vertextyp3/.style={color=graphic3, shape=diamond, vertex}}
\tikzset{vertextyp4/.style={color=graphic4, regular polygon, regular polygon sides=5, vertex}}
\tikzset{vertextyp5/.style={color=graphic5, regular polygon, regular polygon sides=8, vertex}}

\tikzset{>=latex'}

\tikzset{label/.style={text=black, node distance=0}}
\tikzset{edge/.style = {->, draw=gray, line width=0.25pt}}
\tikzset{edgestrong/.style = {edge, line width=1pt}}
\tikzset{tag/.style = {pos=0.5, sloped, fill=White, outer sep=5pt}}

\tikzset{rotation/.style={fill=white, shape=circle, inner sep=0pt}}

\newcommand{\tikzpath}{./}
\newcommand{\pdfpath}{./}

\iftikz\tikzset{external/system call={/usr/local/texlive/2015/bin/x86_64-linux/lualatex -shell-escape -halt-on-error -interaction=batchmode -jobname "\image" "\texsource"}}\fi

\newcounter{Fig}

\newcommand{\includetikz}[1]{\iftikz\begin{minipage}[b]{\figwidth}\vspace{0pt}\scriptsize\centering\tikzincexp\tikzsetnextfilename{\tikznamee{#1}}\input{\tikzpath/#1.tikz}\end{minipage}\else\includetikzpdf{#1}\fi}
\newcommand{\includetikznb}[1]{\iftikz\tikzincexp\tikzsetnextfilename{\tikznamee{#1}}\input{\tikzpath/#1.tikz}\else\includetikzpdf{#1}\fi}
\newcommand{\includetikzex}[1]{\iftikz\includetikzextkz{#1}\else\includetikzexpdf{#1}\fi}

\newcommand{\tikznameexp}[1]{#1\tikzsuf}
\newcommand{\tikznameseq}[1]{\tikzbase\theFig\tikzsuf}

\newcommand{\includetikzpdf}[1]{\centering\tikzincexp\includegraphics{\pdfpath/\tikznamee{#1}.pdf}}
\newcommand{\includetikzextkz}[1]{\tikzincseq\tikzsetnextfilename{\tikznames{#1}}{#1}}
\newcommand{\includetikzexpdf}[1]{\tikzincseq\includegraphics{\pdfpath/\tikznames{#1}.pdf}}
\newcommand{\figwidth}{\columnwidth}
\newcommand{\tikzwidth}[1]{\renewcommand{\figwidth}{#1}}

\newcommand{\tikzincseq}{\stepcounter{Fig}}
\newcommand{\tikzincexp}{}

\newcommand{\tikzbase}{FigLegend}
\newcommand{\tikzsuf}{}
\newcommand{\tikznamee}[1]{\tikznameexp{#1}}
\newcommand{\tikznames}[1]{\tikznameseq{#1}}

\colorlet{graphic1s}{NavyBlue}
\colorlet{graphic2s}{BrickRed}
\colorlet{graphic3s}{Goldenrod}
\colorlet{graphic4s}{ForestGreen}
\colorlet{graphic5s}{Peach}
\colorlet{graphic1}{graphic1s!60}
\colorlet{graphic2}{graphic2s!60}
\colorlet{graphic3}{graphic3s!60}
\colorlet{graphic4}{graphic4s!60}
\colorlet{graphic5}{graphic5s!60}
\colorlet{graphichighl}{Red}
\colorlet{torus}{Grey!80!White}
\pgfplotsset{colormap={graphicmap}{color=(graphic1s) color=(graphic3s) color=(graphic2s)}}

\newcommand{\showscatter}[1]{\raisebox{1pt}{\includetikzex{\tikz \node[vertexcol=#1, minimum width=\widthsm, minimum height=\widthsm] at (0,0) {};}}}

\floatstyle{plain}
\newfloat{si}{ht}{los}
